\documentclass[12pt]{article}

\title{Non stationary nucleation: the model  with averaged velocity}
\author{Victor Kurasov}
\date{St.Petersburg State University}

\begin{document}
\maketitle

\begin{abstract}

A new model to calculate the rate of nucleation is formulated. This model is based on the classical nucleation theory but considers also vapor depletion around the formed embryo. The key characteristic which arises in frames of this theory is the mean time of the embryo formation. On the base of this time the probability to observe the embryo formed during the given time is estimated which allows to construct a simple approximate theory.

\end{abstract}

\section*{Introduction}

This publication deals with phenomena of nucleation and can be regarded as an extension or may be as an alternative approach for the theory presented in \cite{1}. All details can be found in \cite{1} and in corresponding bibliographic references in \cite{11}. Recall that the start of systematic investigations was given already at the end of the 19-th century \cite{Wilson}. The central model adopted in nucleation is the  "classical theory of nucleation" which is the set of assumptions and theoretical constructions presented in papers of  Becker and  Doering \cite{BD}, Volmer and Weber \cite{Volmer}, Zeldovich \cite{Zeldovich}, Frenkel \cite{Frenkel}.

The most solid based  versions of improvements  of the classical theory of nucleation are published by Lothe and Pound \cite{LP} and by  H. Reiss, J. L. Katz, E.R. Cohen \cite{RKC} but these modifications can not also bring the theory in coincidence with experiment.

Now we specify the matter of discussion.
One has to stress that the difficulty for the theoretical description lies not in the form of the rate of nucleation which is
$$
J_{theor} = N_{eq} Z exp(-F_c)
$$
where $F_c$ is a free energy of the critical embryo, $N_{eq}$ is the normalizing factor of the equilibrium distribution, $Z$ is so-called Zeldovich factor. This form is practically evident. The main technical efforts of the classical nucleation theory were devoted to  determine the value of $Z$. But the problem is mainly not in determination of $Z$ which is evidently proportional to the kinetic coefficient $W_+$ (which is the inverse mean time between collisions of the critical embryo with molecules of the condensation substance in vapor) and to the condensation coefficient $\alpha_c$. One can make a mistake for $Z$ which results in some essential error, but not in orders of this magnitude. So, the error seems to be made in $N_{eq}$ of in $F_c$. Since these values appeared together in the value of the equilibrium distribution  $N_{eq} \exp(-F_c)/ \exp(1)$ at the left boundary of the near-critical region one can not separate them (here $\exp(-1)$ appeared only due to specific choice of the left boundary of the near-critical region where the free energy $F$ is prescribed to be $F_c-1$). Historically it is preferable to speak about the error in $F_c$ regarding $N_{eq}$ as some practically constant value like the number of monomers in a system.

Many attempts to improve the classical theory did not lead to the total success. First of all one has to stress the improvement made by Courtney \cite{Courtney} who insisted that it is necessary to divide the flow (the rate of nucleation) by the value of the supersaturation. This approach was developed by Katz and  Weidersich \cite{KW} who multiplied the rate on exponent of the normalized surface tension in addition to the already existed correction. This correction seems to be natural when we recall that the formal continuation of the free energy to the embryo with one molecule leads to the factor of  Katz and  Weidersich. If we decide that the embryo with one molecule has no surface tension we come to the factor of  Courtney. But the problem is that all these approaches did not explain all experimental data.

The last investigation by Thomas P. Bennett and Jonathan C. Barrett \cite{BB} also demonstrates the absence of precise coincidence between the theoretical and experimental results. The relevant citations of modern approaches can be found there.

Despite the universal form of the final stage of evolution \cite{LS}, \cite{Mygolden}, there are many arguments showing that the evolution at considerable times strongly depends on initial position \cite{Mylate}, \cite{Mylate2}, \cite{Myperturbative}, i.e. namely on the effects of the embryos formation. This demonstrates the necessity to improve the theoretical description of the nucleation phenomenon.

 \section{Necessity of a new approach}

What theoretical reasons can one suggest for the necessity of non-stationary approach? It seems that it is enough well grounded that the embryo is the equilibrium object which allows to use the conception of a minimal work to calculate the free energy of the embryo formation. Recall these reasons.

The characteristic time of the embryo relaxation $t_{init}$ can be found from the heat conductivity equation and looks like
$$
t_{init} = (R_d^2/4 D_t)^{1/2}
$$
where $D_t$ is the thermal conductivity coefficient, $R_d$ is the radius of the embryo.

The characteristic time of the embryo perturbation  $t_{coll}$ can be estimated as the inverse flow of the molecules of the condensed substance on the embryo
$$
t_{coll} = W_+^{-1}
$$
The flow of the molecules on the embryo can be easily calculated in the free molecular regime using the gas kinetic theory
$$
W_+ = \frac{1}{4} n v_t S
$$
Here $n$ is the density of the number of condensed substance molecules in vapor, $v_t$ is the mean thermal velocity of these molecules, $S$ is the surface square of the embryo.

In all other regimes of the substance exchange between the embryo and the environment one can prove an estimate
$$
W_+ \leq \frac{1}{4} n v_t S
$$
where $n$ is the density far from the embryo and the embryo is supposed to be the growing one.

Since the volume per one molecule of the condensed substance $v_v$ in vapor is many times greater than the volume of the molecule in the liquid phase $v_l$ (i.e. in the embryo) one can easily see that
$$
t_{init} \ll t_{coll}
$$
Namely this inequality traditionally justifies the applicability of the minimal work for calculation of the equilibrium distribution of embryos.

In the very dense phases (both mother and the new ones) it is possible to observe regimes with the absence of quasi-stationary state of the substance inside the embryo. Then it is necessary to find the specific stationary regimes for the description of the regular evolution. For the fluctuational evolution the problem is more difficult and this forms the content of the special consideration.

Certainly, here we assume that the embryo is an equilibrium object, but in what environment it has to be embedded? Ordinary the embryo is embedded in the vapor with  the    particle density $n_{ext}$ which is the uniform density initiated by conditions of experiment.  But the embryo consumes the vapor molecules which leads to some density profile around the embryo. Actually this question is taken into account in the diffusion regime of the embryos growth where the density profile is calculated due to the regular growth while here the fluctuational growth is considered which is assumed to be many times faster.

The density $n$ becomes the function of the distance from the embryo $r$ and it is lower than $n_{ext}$
$$
n(r) \leq n_{ext}
$$

The small values of $n$ near the embryo correspond to the big values of the size of the critical embryo and the small values of the nucleation rate. Here appeared an unpleasant question: at what distances? The distance can not be too small. As it will be shown later, this question will be connected with the number of steps used in approximate description of kinetics.

 One can approximately get the profile $n$ from the Green function of the diffusion equation
$$
n_{ext} - n(r) =
\int_0^t v_{\nu} (t') \frac{1}{\sqrt{4\pi D (t-t')}}
\exp(-\frac{r^2}{4D(t-t')}) dt'
$$
 Here $D$ is the diffusion coefficient in vapor and the value $v_{\nu}$ is the velocity $d\nu/dt$ of the embryos growth. Certainly, it is not a regular value but some realization of the individual story of some embryo. It is not a sum of delta-functions but has to be averaged over some time interval which corresponds to the number of steps in the model (see later).

This approximation is rather good regarding embryo as the point consumer of vapor. The approximation of the point source can be used due to the same strong inequality
$$
v_v\gg v_l
$$

A rough approximation going also from the last formula is the following: when the time from the embryos formation is $t_0$ then the embryo consumes vapor from the vapor region of the linear size
$$
R_0 = \sqrt{4Dt_0}
$$

The natural way to ensure the greatest $n$ (i.e. $n_{ext}$) is to wait as long as possible. The key question of the consideration which arises here sounds as following: Is it possible to wait infinitely long? The answer which lies in the base of the approach which is presented below (and in \cite{1}) is "No".

Really recall that for every embryo there exists a time of the dissolution $t_{diss}$. For the first time it appeared in consideration of the relaxation to stationary distribution. This value can be introduced as
$$
- \nu = \int_0^{t_{diss}} v_{reg}(t') dt'
$$
where $v_{reg}$ is velocity of the regular motion. The last value can be presented as
$$
v_{reg} = W_+  - W_-
$$
where
$W_-$ is the inverse flow, i.e. the flow of molecules
from the embryo to the vapor.

Since both $W_+$ and $W_-$ are originally the functions of the embryos size $\nu$ it is better to rewrite the expression for the dissolution time as
$$
t_{diss} = \int_0^{\nu} \frac{1}{|W_+(\nu') - W_-(\nu')|} d\nu'
$$

Certainly. the last formula can not be applied in the neighborhood of the critical point where one has to consider diffusion, but this region is rather small.

For us the significance of this time is clear - the embryo can not exist essentially longer than $t_{diss}$ in the pre-critical region (here $\nu$ is some characteristic size from this region). Certainly, precisely speaking  this conclusion is wrong, the evolution occurs due to fluctuations, but one can see that the probability to stay long in the pre-critical region is low.

The main conclusion which comes from the arguments presented above is that one has to take into account that the embryo can not grow up to the critical value infinitely long. One can show that the probability for the embryo to grow up to the critical size during some time $t$  will go to zero when $t$ is relatively big.

 To construct the theory with explicit formulas one has to determine the probability for the embryo to grow up to the critical size during some time $t$.  Certainly, it is rater difficult to suggest a suitable expression for such probability, but some approximation will be given.

Now it is necessary to see the difference between this approach and the approach given in \cite{1}. In \cite{1} instead of presenting a reliable expression for the mentioned probability it was assumed that the faster the embryo is formed the greater is not only the probability but also the value of the probability  divided by exponent of the embryos free energy, i.e. total rate of nucleation. Certainly, there are some strong arguments for such conclusion.  These arguments were presented in \cite{1}.

How one can act under the situation studied in \cite{1}? The necessary restriction which allows to construct the theory came from the requirement of stability of the system. Fortunately there exists a limit of stability for the system of an embryo with environment and this limit was taken in \cite{1} as a true characteristic for the embryo going to be a critical one.

Meanwhile the assumption of the minimal environment around the embryo ensuring the stability of this embryo looks very attractive it would be interesting to present the explicit expression for the probability to stay in a fixed region during a given time. It allows to construct the explicit theory and determine the rate of nucleation. This rate of nucleation will differ from the value given in \cite{1}. Namely this program is fulfilled below.

\section{Estimate for probability}

What estimate for the embryo to stay in the pre-critical region one can suggest? Here we use the simplest variant based on the Green function. If the regular velocity of the embryo is $-v_r$ and it is supposed to be constant at least in the region of essential localization of the Green function then the probability for the embryo which had $\nu=x_0$ at $t_0$ to have $\nu$ in interval $[x,x+dx]$ at time $t$ is given by $G(x,t|x_0,t_0)dx$ where
$$
G(x,t|x_0, t_0) =
\frac{1}{\sqrt{4\pi D (t-t_0)}}
\exp(-\frac{(x-x_0+v_r (t-t_0))^2}{4 D (t-t_0)})
$$

So, $G$ is simply a compressed and shifted gaussian.

It is extremely important that here we see the same form that appeared in the law of large numbers and in the central limit theorem in mathematical statistics. This allows to reformulate all constructions in frames of stochastic events and then to use instruments of mathematical statistics including the interval estimation.

For us the point of interest will be the value at the tail of $G$.  We shall say according to the interval approach in mathematical statistics
that the probability for embryo not to loose the molecules is equal to the integral over the tail of the Green's function, i.e.
$$
P_0 = \int_{x_0}^{\infty}
\frac{1}{\sqrt{4\pi D (t-t_0)}}
\exp(-\frac{(x-x_0+v_r (t-t_0))^2}{4 D (t-t_0)}) dx
$$

Here appears the evident disagreement because when $t$ goes to $t_0$ then the probability $P_0$ goes to $1/2$ instead of $1$. So, it has to be multiplied on $k \approx 2$ and the last expression looks like
  $$
P_0 = \int_{x_0}^{\infty}
\frac{k}{\sqrt{4\pi D (t-t_0)}}
\exp(-\frac{(x-x_0+v_r (t-t_0))^2}{4 D (t-t_0)}) dx
$$
The possible explanation is the  following  one - approximately the probabilities to go faster than the mean velocity and slower then the mean velocity are the same and integrating over the tail we take into account only one of them. For small times both these possibilities lead to the evident result - to stay near the initial value. So, we have to take both of them.

The above mentioned arguments are valid for small $t-t'$. For big $t-t'$ the motivation is another. Since we integrate over the whole length of the  tail we prescribe the conservation of the size for all embryos from the tail. The characteristic width of this distribution will be
$$
\Delta = (\frac{d(\frac{(x-x_0+v_r (t-t_0))^2}{4 D (t-t_0)})}{dx})^{-1}
$$
The derivative here is taken at $x=x_0$
The same resolution has to be applied for the embryos which do not attain $x$ which will lead to
  $$
 \int_{x_0-\Delta}^{\infty}
\frac{1}{\sqrt{4\pi D (t-t_0)}}
\exp(-\frac{(x-x_0+v_r (t-t_0))^2}{4 D (t-t_0)}) dx
$$
and approximately gives the already presented formula.

Certainly, here we meet the weakest point of this approach. But if it will be shown that for some $k$ from
the interval
$[1,2]$ there is an approximate coincidence with experiments one can refine the model.

One can state even more: In further final conclusions the concrete value of the constant parameter $k$ is not important. It drops out from formulas for the rate of nucleation.

The next problem is how to calculate the integral. For our purposes the interesting situation will be the situation when there is a small tail and the integral can be approximately taken as
$$
P_0 = \int_0^{\infty}
\frac{k}{\sqrt{4\pi D (t-t_0)}}
\exp(-\frac{(v_r (t-t_0))^2}{4 D (t-t_0)}) \exp(-\Delta y) dy
$$

The last integral can be easily taken which gives
$$
P_0 =
\frac{k}{\sqrt{4\pi D (t-t_0)}}
\exp(-\frac{(v_r (t-t_0))^2}{4 D (t-t_0)}) / \Delta
$$

All these calculations will be useful below, here we are interested in these manipulations only to see  that the probability for the embryo to stay infinitely long is infinitely small. So, one can see that the overcoming of the activation barrier occurs in a strongly non-stationary manner. This requires to take the mentioned fact into account in construction of the theory.

To complete the calculation of the integral which will be necessary below we investigate other characteristic situations.
In any vase the integral can be precisely reduced to the error function and for the error function one can use the well known Boyd's  approximation
$$
 \frac{\pi/2}{\sqrt{z^2+ \pi} + (\pi-1) z}  \leq \exp(z^2) \int_z^\infty \exp(-t^2) dt \leq
\frac{\pi/2}{\sqrt{(\pi-2)^2 z^2+\pi}+2z}
$$
for $z>0$. Precision of this approximation is relatively high.

The previous variant of the non-stationary description \cite{1} was very approximate one. The only property of nucleation which was required in \cite{1} was the relative stability of cluster. This model does not regard the  minimal environment which is necessary to provide the stable cluster. Certainly, if the cluster will be formed at the longer period and from the larger environment then the free energy of the self-formation (without the properties of the low probability of the slow formation of an embryo) will be lower. So, it is worth to describe this situation. At least one has to take into account explicitly the time of the embryo formation. It will be done below.

Since the condensation of one molecule in a liquid phase leads to the heat extraction one has to take
into account the thermal effects. Since the thermal conductivity equation and diffusion equation have one and the same form one can omit here (in the simplest variant of the theory to grasp the idea of this approach) the thermal effects having assumed that they can be taken into account by the scale transformations.

\section{The model}

Precisely speaking one has to consider all trajectories of motion for the embryo, then to construct the Green functions with corresponding intensity of the vapor consumption and then to integrate over all possible trajectories multiplied on probabilities for the embryo not to be dissolved moving along this trajectory (i.e. the probability to take this trajectory). This procedure is too complex and we shall present here the simplest approximate variant.

The main object of interest is the critical embryo (here it differs from the equilibrium critical embryo). When it is formed it will grow practically irreversibly (the overcoming of the near-critical region can lead to some corrections of the order of the half-width of the near-critical region, but here we search for the quantities of the leading order). Since the region of environment enlarges, the density grows and the embryo becomes to be the supercritical one automatically without moving along $\nu$-axis. Later all characteristics of this embryo will be determined. Suppose that the time of it's formation is $T$. Then the perturbation from creation of the embryo will be spread over the distance
$$
R_{res} = \sqrt{4 D T }
$$
More precisely one can construct the Green functions solution in the following form
$$
n_{ext} - n =
\int_0^t v_f(t') \frac{1}{\sqrt{4 \pi D(t-t')}}
\exp(-\frac{r^2}{4D(t-t')}) dt'
$$
where $v_f$ is the velocity of the averaged fluctuational embryos growth. It is reasonable to take as $v_f$ the value
$$
v_f = \nu_c / T
$$
and then one can calculate the last integral in terms of the error function.   The result will be the same: $R_{res}$ is the characteristic radius of the environment of perturbation. Namely from this environment the vapor consumption can occur. Then the average density decreases. The value $n_{av}$ of the reduced density in this region will be
calculated on the base of the following relation
$$
(n_{ext} - n_{av} ) \frac{4}{3} \pi R_{res}^3 = \nu_c
$$
or
$$
n_{av} = n_{ext} - \frac{3}{4\pi} \nu_c R_{res}^{-3}
$$
and finally
$$
n_{av} = n_{ext} - \frac{3}{4\pi} \nu_c (4 D T)^{-3/2}
$$

Namely, on the base of $n_{av}$ the free energy of the embryo has to be constructed
$$
F = - \ln(n_{av}/n_{\infty}) \nu + a \nu^{2/3}
$$
which gives
$$
F_c = \frac{a}{3} \nu_c^{2/3}
$$
$$
\nu_c = (\frac{2a}{3 \ln(n_{av}/n_{\infty})})^3
$$

The flow or the pure rate of nucleation will look like
$$
J = N^{tot} \exp(-\frac{a}{3} (\frac{2a}{3 \ln(n_{av}/n_{\infty})})^2) Z
$$
where $N^{tot}$ is a normalizing factor.

The total rate is the pure rate multiplied on the probability $P_0(T)$ for the embryo to wait the time $T$. This value is
  $$
P_0 (T) = \int_{x_0}^{\infty}
\frac{k}{\sqrt{4\pi D T}}
\exp(-\frac{(x-x_0+v_r T)^2}{4 D T}) dx
$$
where $v_r$ is the averaged velocity.

 Then we shall search for the minimum of the function
 $$
 \Psi = \exp(-\frac{a}{3} (\frac{2a}{3 \ln(n_{av}/n_{\infty})})^2)
 \int_{x_0}^{\infty}
\frac{k}{\sqrt{4\pi D T}}
\exp(-\frac{(x-x_0+v_r T)^2}{4 D T}) dx
$$
or
 $$
 \Psi = \exp(-\frac{a}{3} (\frac{2a}{3 \ln((n_{ext} - \frac{3}{4\pi} \nu_c (4 D T)^{-3/2})/n_{\infty})})^2)
 \int_{x_0}^{\infty}
\frac{k}{\sqrt{4\pi D T}}
\exp(-\frac{(x-x_0+v_r T)^2}{4 D T}) dx
$$

Here one has to put $v_r$ as
$$
v_r =
W_+ [\exp(-\frac{F_c}{\nu_c})-1]
\approx
W_{+c} [\exp(-\frac{F_c}{\nu_c})-1]
$$
or
$$
v_r
=
W_{+c} [ \exp(-(a/3)\nu_c^{-1/3}(T)) -1]
$$

 This gives
$$
v_r
=
W_{+c0} \nu_c^{2/3} [\exp(-(a/3)\nu_c^{-1/3}(T))-1]
$$
where $W_{+c0}$ is the characteristic constant to extract the explicit dependence on $\nu_c$.

Since $W_+$ 9and $W_-$) is proportional to $\nu^{2/3}$ it is worth to rewrite the theory in terms of the new variable where this dependence is absent. It is quite easy to do. 
Having noted that
 $$
\nu_c(T) = (\frac{3 \ln((n_{ext} - \frac{3}{4\pi} \nu_c (4 D T)^{-3/2})/n_{\infty})}{2a})^{-3}
$$
we get the closed equation on $T$ which can be reduced with the help of the mentioned approximations to algebraic equation on $T$. This equation is so simple that it can be solved with the help of elementary methods. This gives the function $\Psi$ as a function of one parameter $T$. We seek for the maximum of $\Psi$. Since the integral in expression for $\Psi$ can be approximately expressed in elementary functions, we have an algebraic expression which can be easily differentiated. The zero of the derivative gives the equation on $T$. This equation is the algebraic one and can be easily solved. Now we know $T$, then we get the rate of nucleation. The task is solved.

It is better to express $T$ via $n_{av}$ since the interval for $n_{av}$ is finite. It is
$[n_{min}, n_{ext}]$ where $n_{min}$ is the value of density calculated in frames of the model of the minimal environment \cite{1}. Then the task to find the root of the algebraic equation at the finite interval is rather simple.

If there are no roots in the mentioned interval it means that it is necessary to use the model with the minimal environment.

\section{The simplest version}

Now we present the simplest version of solution of algebraic equations based on one hand on decomposition not far from the classical case. But nevertheless the effect is supposed to be essential. Namely in this case one can fulfill very simple analysis of the equation on $T$.

For $P$ one can write
$$
P =
\frac{k}{\sqrt{4 \pi DT}}
\int_0^{\infty} \exp(-(y+v_rT)^2 / 4 DT) dy \equiv \frac{k}{\sqrt{4 \pi DT}}
I
$$
For essential deviations of the flow from the classical expression is it necessary that $P \ll 1$. This leads to
$$
 \exp(-(v_rT)^2 / 4 DT) \ll 1
 $$
Then
 $$
I
 \approx
 \exp(-(v_rT)^2 / 4 DT) \int_0^{\infty}
 \exp(-2 y v_r T/ (4DT)) dy
$$
 and
$$
I
\approx
 \exp(-(v_rT)^2 / 4 DT) \frac{1}{2  v_r / (4D)}
 $$

  In further constructions one can put $\nu_c$ as a constant because
  $
  \exp(-\frac{a}{3}\nu_c^{2/3}) $ for moderate $a/3$  changes essentially when the relative variation in $\nu_c$ is small (here $a$ is moderate).

  Then the expression for $\Psi$ can be approximately rewritten as
  $$
  \Psi =
  \exp(-\frac{a}{3} (\frac{2a}{3 \ln((n_{ext} - \frac{3}{4\pi} \nu_c (4 D T)^{-3/2})/n_{\infty})})^2)
  \frac{k}{\sqrt{4 \pi DT}} \exp(-(v_rT)^2 / 4 DT) \frac{1}{2  v_r / (4D)}
  $$
  Instead of  extremum for $\Psi$ we can approximately consider extremum for
  $$
  \phi =\frac{a}{3} (\frac{2a}{3 \ln((n_{ext} - \frac{3}{4\pi} \nu_c (4 D T)^{-3/2})/n_{\infty})})^2
 +v_r^2 T / 4 D
 $$
 For small deviations of
 $$
 \nu_c^{1/3} = \frac{2a}{3 \ln((n_{ext} - \frac{3}{4\pi} \nu_c (4 D T)^{-3/2})/n_{\infty})}
 $$
 we have the following expression for $T$
 $$
T=(\nu_c^2 v_r^{-2} \frac{9}{8\pi} \frac{1}{n_{ext}} (4D)^{-1/2})^{2/5}
$$

\section{Discussion}

Certainly this model is only the first step in construction of the realistic models corresponding to the non-stationary formation of the critical embryo and to the non-stationary rate of nucleation. It is very difficult to perform the complete task to investigate all possible temporal trajectories to overcome the barrier, then to calculate the probability to form an embryo at this trajectory taking into account the depletion of the mother phase around the embryo. So, this model has to be regarded as an instrument to see whether the experimental tendencies of deviation from the theoretical formulas can be explained by the non-stationary effects of the critical embryo formation.

There exists an attractive possibility to split the interval $[1,\nu_c]$ into several steps, then to consider the sequential overcoming of all steps up to the critical value in a manner as it is done above. Here all constructions will be absolutely analogous to the already written ones except the necessity to account of depletion from all previous steps. It is easy to do, no principal obstacles will appear, the only difficulty is to write some simple but huge formulas. One can even write the differential analog of this process and  solve the variational problem. It is also possible to do analytically while there will be necessary to use some approximative transformations.

The difficulty of splitting the process in many steps lies in another field - in the applicability of the model for such small time intervals. On one hand one can not consider here the diffusion equation and any profile of density. On the other hand the emission or ejection of one molecule at such small intervals causes the jump of density to the magnitudes corresponding to the unstable state and can not be applied.

There exists one more specific feature in consideration of small intervals - the problem of reverse transitions. When we consider large intervals there is no probability that the embryo come back, then continue to grow and reach the critical value. But for  small intervals this probability exists. When
 the probability is small it can be taken into account by summation like the sum of some geometric progressions, but when it is essential the trajectory can be very complex and one can not fulfill calculation only by summation of the geometric progressions. So, here the problem becomes very complex.

One can not say that the reverse movements are absent in one-step consideration. They are presented in the gaussian form of the Green-function of the diffusion equation. Then for two-step consideration with reverse transitions one has to fulfil the convolution of probabilities. For two gaussian curves  this convolution gives again a gaussian but here it is necessary to account the non-gaussian factors associated with the free energies and this makes impossible to fulfill this convolution analytically. Only the case of small corrections to triangle approximation can be investigated in frames of the perturbation theory.

So, here the model of the non-stationary nucleation is formulated.
One can use here instead of the classical theory of nucleation all other already existing theories including the density functional approach. So, it is possible to speak about a new family of the nucleation theories. The current task is to check the coincidence between the theoretical and experimental data.

  \end{document}